\documentclass[11pt]{cernrep}
\usepackage{graphicx}
\usepackage{here}
\begin{document}
 \title{Confinement, Chiral Symmetry and Hadrons}
\author{\'Agnes M\'ocsy$^{a,b}$, Francesco Sannino$^{b}$\thanks{Speaker at the
workshop.} and Kimmo Tuominen$^{b}$}  \institute{$^a$ Institut
f\"ur Theoretische Physik, J.W.~Goethe Universit\"at, Postfach
111932, 10054 Frankfurt am Main, Germany
\\$^b$ The Niels Bohr Institute \& NORDITA, Blegdamsvej 17, Copenhagen DK-2100, Denmark.}
 \maketitle
\begin{abstract}
The Polyakov loop is the appropriate deconfinement order parameter
for Yang-Mills theories without quarks or with quarks in the
adjoint representation of the gauge group. However it is not a
physical state of the theory so the information regarding the
center group symmetry must be transferred to the physical states.
We briefly review how this transfer of information takes place.
When adding quarks the center group is no longer a symmetry for
matter in the fundamental representation. This feature allows us
to explain why color deconfines when chiral symmetry is restored
in hot gauge theories with massless quarks. {}For quarks in the
adjoint representation we show that while deconfinement and the
chiral transition do not coincide, entanglement between them is
still present. Finally, we discuss also the chemical potential
driven phase transition.
\end{abstract}


Recently in \cite{Sannino:2002wb,{Mocsy:2003tr},{Mocsy:2003qw}} we
have analyzed the problem of how, and to what extent the
information encoded in the order parameter of a generic theory is
transferred to the non-order parameter fields. This is a
fundamental problem since in nature most fields are non-order
parameter fields. This problem is especially relevant in QCD and
QCD-like theories since there is no physical observable for
deconfinement which is directly linked to the order parameter
field. At nonzero temperature the $Z_{N_c}$ center of $SU(N_c)$ is
a relevant global symmetry \cite{Svetitsky:1982gs}, and it is
possible to construct the Polyakov loop $\ell$ operator.
This object is charged with respect to the center $Z_{N_c}$ of the
$SU(N_c))$ gauge group \cite{Svetitsky:1982gs}, under which it
transforms as $\ell \rightarrow z \ell$ with $z\in Z_{N_c}$. A
relevant feature of the Polyakov loop is that its expectation
value vanishes in the low temperature regime, and is non-zero in
the high temperature phase. The Polyakov loop is thus a suitable
order parameter for the Yang-Mills temperature driven phase
transition \cite{Svetitsky:1982gs}.

The question is: How can the information about the Yang-Mills
phase transition encoded in the $Z_{N_c}$ global symmetry be
communicated to the hadronic states of the theory. In
\cite{Sannino:2002wb,{Mocsy:2003tr}}, using an effective
Lagrangian approach, we have explicitly shown that is always
possible on symmetry grounds to couple $\ell$ to the hadronic
operators. Effective Lagrangians play a relevant role. In
\cite{Sannino:2003xe}, for example, a number of fundamental
properties for QCD were derived. Of course, at a more elementary
level the picture is that the gauge invariant operators of the
theory carrying center group charges couple to hadronic states. If
not protected by symmetries only an unnatural act would require
these relevant couplings to vanish. Such a picture is also
confirmed within theoretical models \cite{Meisinger:2002ji}. We
have also showed \cite{Mocsy:2003tr} that spatial correlators of
the non-critical field are dominated at the phase transition by
the critical behavior of the order parameter field. In this way
the Polyakov loop field imprints the knowledge of the phase
transition on the non order parameter spatial correlators (e.g.
the glueball field). Our results are applicable to any phase
transition and as such are universal.

The picture changes considerably when quarks are added to the
theory. If fermions are in the fundamental and pseudoreal
representations for $N_{\rm c}=3$ and $N_{\rm c}=2$, respectively,
the corresponding $Z_3$ or $Z_2$ center of the group is never a
good symmetry. {}For massless quarks the order parameter is the
chiral condensate which characterizes the chiral phase transition.
{}For $N_c$=3 and two massless quark flavors at finite temperature
and zero baryon density, the chiral phase transition is in the
same universality class as the three dimensional $O(4)$ spin model
\cite{Wilczek:1992sf}, becoming a smooth crossover as small quark
masses are accounted for \cite{Scavenius:2000qd}. {}For $N_c=2$
the relevant universality class is that of $O(6)$ both for the
fundamental and adjoint representations \cite{Holtmann:2003he}.
Even if the discrete symmetry is broken, one can still construct
the Polyakov loop and study the temperature dependence of its
properties on the lattice. One still observes a rise of the
Polyakov loop from low to high temperatures and naturally,
although improperly, one speaks of deconfining phase transition
\cite{Karsch:1998qj}. {}For fermions in the adjoint representation
the center of the group remains a symmetry of the theory, and thus
besides the chiral condensate, also the Polyakov loop is an order
parameter.

Interestingly, lattice results \cite{Karsch:1998qj} indicate that
for ordinary QCD with quarks in the fundamental representation,
chiral symmetry breaking and confinement (i.e. a decrease of the
Polyakov loop) occur at the same critical temperature. Lattice
simulations also indicate that these two transitions do not happen
simultaneously when the quarks are in the adjoint representation.
Despite the attempts to explain these behaviors \cite{Brown:dm},
the underlying reasons are still unknown. Recently, using an
interesting model containing only hadrons this issue has also been
addressed for ordinary QCD in \cite{Agasian:2003ux}.

In \cite{Mocsy:2003qw} we have proposed a solution to this puzzle
based on the general approach established in \cite{Mocsy:2003tr},
envisioned first in \cite{Sannino:2002wb}. Two general features
introduced in \cite{Mocsy:2003tr} are essential: There exists a
relevant trilinear interaction between the light order parameter
and the heavy non-order parameter field, singlet under the
symmetries of the order parameter field. This allows for an
efficient transfer of information from the order parameter to the
fields that are singlets with respect to the symmetry of the
theory. As a result, the non-critical fields have infrared
dominated spatial correlators. The second feature, also due to the
existence of such an interaction, is that the finite expectation
value of the order parameter field in the symmetry broken phase
induces a variation in the expectation value for the singlet
field, whose value generally is non-vanishing in the unbroken
phase.

\section{Fundamental Representation}
\label{fundamental}

Here we study the behavior of the Polyakov loop by treating it as
a heavy field that is a singlet under chiral symmetry
transformations. We take the underlying theory to be two colors
and two flavors in the fundamental representation. The degrees of
freedom in the chiral sector of the effective theory are
$2N_f^2-N_f-1$ Goldstone fields $\pi^a$ and a scalar field
$\sigma$. {}For $N_f=2$ the potential is \cite{Appelquist:1999dq}:
\begin{eqnarray}
V_{\rm ch}[\sigma,\pi^a]&=&\frac{m^2}{2}{\rm Tr
}\left[M^{\dagger}M\right]+ {\lambda_1}{\rm Tr
}\left[M^{\dagger}M\right]^2+ \frac{\lambda_2}{4}{\rm Tr
}\left[M^{\dagger}MM^{\dagger}M\right] \label{chiralpot}
\end{eqnarray}
with $2\,M=\sigma + i\,2\sqrt{2}\pi^a\,X^a$, $a=1,\dots,5$ and
$X^a\in {\cal A}(SU(4))-{\cal A}(Sp(4))$. $X^a$ are the generators
provided explicitly in equation (A.5) and (A.6) of
\cite{Appelquist:1999dq}. The Polyakov loop potential in the
absence of the $Z_2$ symmetry is
\begin{eqnarray}
V_\chi[\chi]=g_0\chi+\frac{m_\chi^2}{2}\chi^2+\frac{g_3}{3}\chi^3
+\frac{g_4}{4}\chi^4 \, . \label{chipot}
\end{eqnarray}
The field $\chi$ represents the Polyakov loop itself. To complete
the effective theory we introduce interaction terms allowed by the
chiral symmetry
\begin{eqnarray}
V_{\rm{int}}[\chi,\sigma,\pi^a]&=& \left(g_1\chi
+g_2\chi^2\right){\rm Tr } \left[M^{\dagger}M\right]=\left(g_1\chi
+g_2\chi^2\right)(\sigma^2+\pi^a\pi^a) \, .
\end{eqnarray}
In the phase with $T<T_{c\sigma}$, where chiral symmetry is
spontaneously broken, $\sigma$ acquires a nonzero expectation
value, which in turn induces a modification also for
$\langle\chi\rangle$. The usual choice for vacuum alinement is in
the $\sigma$ direction, i.e. $\langle\pi\rangle=0$. 
The extremum of the linearized potential is at
\begin{eqnarray}
\langle\sigma\rangle^2 &\simeq& -\frac{m^2_{\sigma}}{\lambda}\ ,
\quad  \quad m^2_{\sigma} \simeq m^2 + 2g_1\langle \chi\rangle
\label{vevsigma} , \quad  {\rm and} \quad\langle \chi \rangle
\simeq \chi_0 -\frac{g_1}{m_\chi^2}\langle\sigma\rangle^2\, \quad
~ \chi_0 \simeq -\frac{g_0}{m^2_{\chi}} \ ,
 \label{vevchi}
\end{eqnarray}
where $\lambda=\lambda_1 + \lambda_2$.  Here $m^2_{\sigma}$ is the
full coefficient of the $\sigma^2$ term in the tree-level
Lagrangian which, due to the coupling between $\chi$ and $\sigma$,
also depends on $\langle\chi\rangle$. Spontaneous chiral symmetry
breaking appears for $m^2_{\sigma} <0$. In this regime the
positive mass squared of the $\sigma$ is $M^2_{\sigma} = 2
\lambda\langle \sigma ^2 \rangle$. The formulae in
(\ref{vevsigma}) hold near the phase transition where $\langle
\sigma \rangle$ is small. We have ordered the couplings such that
$g_0/m_{\chi}^3$ and $g_1/m_{\chi}$ are both much greater than $
g_2$ and $g_3/m_{\chi}$.
%
Near the critical temperature the mass of the order parameter
field is assumed to posses
the generic behavior $m_\sigma^2\sim (T-T_{\rm{c}})^\nu$. Equation
(\ref{vevchi}) shows that for $g_1>0$ and $g_0<0$ the expectation
value of $\chi$ behaves oppositely to that of $\sigma~$: As the
chiral condensate starts to decrease towards chiral symmetry
restoration, the expectation value of the Polyakov loop starts to
increase, signaling the onset of deconfinement. This is
illustrated in the left panel of figure \ref{Figura1}.

When applying the analysis presented in \cite{Mocsy:2003tr}, the
general behavior of the spatial two-point correlator of the
Polyakov loop can be obtained. Near the transition point, in the
broken phase, the $\chi$ two-point function is dominated by the
infrared divergent $\sigma$-loop. This is so, because the $\pi^a$
Goldstone fields couple only derivatively to $\chi$, and thus
decouple. We find a drop in the screening mass of the Polyakov
loop at the phase transition. When approaching the transition from
the unbroken phase the Goldstone fields do not decouple, but
follow the $\sigma$, resulting again in the drop of the screening
mass of the Polyakov loop close to the phase transition. We
consider the variation of the $\chi$ mass near the phase
transition with respect to the tree level mass $m_{\chi}$ defined
above the chiral phase transition. We define $\Delta
m_\chi^2(T)=m_\chi^2(T)-m_\chi^2$, where $m_{\chi}$ is the mass at
a given temperature near the critical point. Using a large $N$
framework motivated resummation \cite{Mocsy:2003tr} we deduce:
\begin{eqnarray}
\Delta m^2_\chi(T)&=& -
\frac{2g_1^2(1+N_\pi)}{8\pi\,m_{\sigma}+(1+N_\pi)3\lambda }
  \quad T>T_{\rm{c}\sigma} \quad {\rm and }
 \quad \Delta m^2_\chi(T)= - \frac{2g_1^2}{8\pi\,M_{\sigma}+ 3\lambda}
  \quad T<T_{\rm{c}\sigma}\, . \nonumber
\end{eqnarray}
From the above equations one finds that the screening mass of the
Polyakov loop is continuous and finite at $T_{\rm{c}\sigma}$, and
$\Delta m_\chi^2(T_{\rm{c}\sigma})=-2g_1^2/(3\lambda)$,
independent of $N_\pi$, the number of pions. This analysis is not
restricted to the chiral/deconfining phase transition. The
entanglement between the order parameter (the chiral condensate)
and the non-order parameter field (the Polyakov loop) is
universal.

\section{Adjoint Representation}
\label{adjoint}

As a second application, consider two color QCD with two massless
Dirac quark flavors in the adjoint representation. Here the global
symmetry is $SU(2N_f)$ which breaks via a bilinear quark
condensate to $O(2N_f)$. The number of Goldstone bosons is
$2N_f^2+N_f-1$. We take $N_f=2$. There are two exact order
parameter fields: the chiral $\sigma$ field and the Polyakov loop
$\chi~$. Since the relevant interaction term $g_1\chi\sigma^2$ is
now forbidden, one might expect no efficient information transfer
between them. This naive statement is supported by lattice data
\cite{Karsch:1998qj}. While respecting general expectations the
following analysis suggests the presence of a new and more
elaborated structure which lattice data can clarify in the near
future.

The chiral part of the potential is given by (\ref{chiralpot})
with $2\,M=\sigma + i\,2\sqrt{2}\pi^a\,X^a$, $a=1,\dots,9$ and
$X^a\in {\cal A}(SU(4))-{\cal A}(O(4))$. $X^a$ are the generators
provided explicitly in equation (A.3) and (A.5) of
\cite{Appelquist:1999dq}. The $Z_2$ symmetric potential for the
Polyakov loop is
\begin{eqnarray}
V_\chi[\chi]=\frac{m_{0\chi}^2}{2}\chi^2+\frac{g_4}{4}\chi^4 \, ,
\end{eqnarray}
and the only interaction term allowed by symmetries is
\begin{equation}
V_{\rm{int}}[\chi,\sigma,\pi]=g_2\chi^2\,{\rm Tr
}\left[M^{\dagger}M\right]
 =g_2\chi^2(\sigma^2+\pi^a\pi^a) \, .
\end{equation}
\begin{figure}[h]
\begin{center} \includegraphics[width=12 truecm,
clip=true]{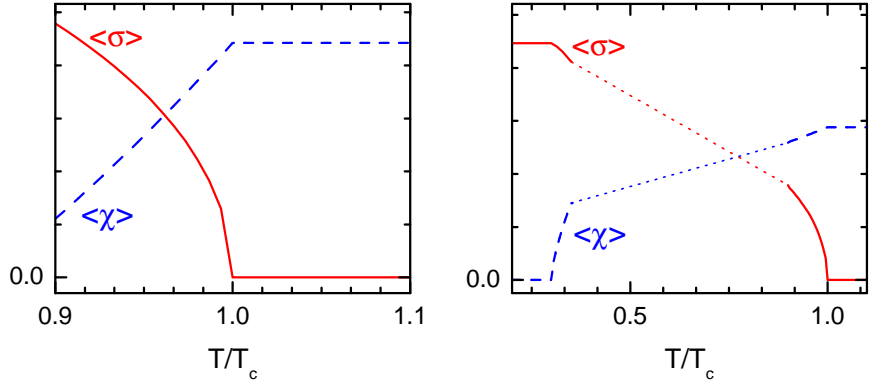}\end{center}\vspace{-1cm}
 \caption{Left panel: Behavior of the expectation
values of the Polyakov loop and chiral condensate close to the
chiral phase transition as a function of the temperature, with
quarks in the fundamental representation. Right panel: Same as in
left panel, for quarks in the adjoint representation and
$T_{\rm{c}\chi}\ll T_{\rm{c}\sigma}$ (see discussion in the
text).} \label{Figura1} \vspace{-.5cm}
\end{figure}
Here we review the case in which the chiral phase transition
happens first, i.e. $T_{{\rm{c}}\chi}\ll T_{{\rm{c}}\sigma}$.
{}For $T_{{\rm{c}}\chi}<T<T_{{\rm{c}}\sigma}$ both symmetries are
broken, and the expectation values of the two order parameter
fields are linked to each other:
\begin{eqnarray}
\langle\sigma\rangle^2&=&-\frac{1}{\lambda}\left(m^2+
2g_2\langle\chi\rangle^2\right)\equiv -\frac{m_\sigma^2}{\lambda}\
, \qquad \langle\chi\rangle^2=-\frac{1}{g_4}\left(m_{0\chi}^2+
2g_2\langle\sigma\rangle^2\right)\equiv -\frac{m_\chi^2}{g_4}\, .
\label{vevad}
\end{eqnarray}
The coupling $g_2$ is taken to be positive. The expected behavior
of $m_\chi^2\sim (T-T_{\rm{c}\chi})^{\nu_{\chi}}$ and
$m_\sigma^2\sim (T-T_{\rm{c}\sigma})^{\nu_{\sigma}}$ near
$T_{\rm{c}\chi}$ and $T_{\rm{c}\sigma}$, respectively, combined
with the result of eq. (\ref{vevad}), yields in the neighborhood
of these two transitions the qualitative situation, illustrated in
the right panel of figure \ref{Figura1}. On both sides of
$T_{\rm{c}\chi}$ the relevant interaction term $g_2\langle
\sigma\rangle\sigma\chi^2$ emerges, leading to a one-loop
contribution to the static two-point function of the $\sigma$
field $\propto \langle \sigma \rangle^2 /m_\chi~$. Near the
deconfinement transition $m_\chi\rightarrow 0$ yielding an
infrared sensitive screening mass for $\sigma$. Similarly, on both
sides of $T_{\rm{c}\sigma}$ the interaction term $\langle
\chi\rangle\chi\sigma^2$ is generated, leading to the infrared
sensitive contribution $\propto \langle \chi\rangle^2/m_\sigma$ to
the $\chi$ two-point function. We conclude, that when
$T_{\rm{c}\chi}\ll T_{\rm{c}\sigma}$, the two order parameter
fields, a priori unrelated, do feel each other near the respective
phase transitions. It is important to emphasize that the effective
theory works only in the vicinity of the two phase transitions.
Interpolation through the intermediate temperature range is shown
by dotted lines in the right panel of figure \ref{Figura1}.
Possible structures here must be determined via first principle
lattice calculations.

The infrared sensitivity leads to a drop in the screening masses
of each field in the neighborhood of the transition of the other,
which becomes critical, namely of the $\sigma$ field close to
$T_{\rm{c}\chi}$, and of the $\chi$ field close to
$T_{\rm{c}\sigma}~$. The resummation procedure outlined in the
previous section predicts again a finite drop for the masses:
\begin{eqnarray}
\Delta m^2_\chi(T_{\rm{c}\sigma})=-\frac{8g_2^2\langle\chi\rangle
^2}{3\lambda}, \qquad \Delta
m^2_{\sigma}(T_{\rm{c}\chi})=-\frac{8g_2^2\langle\sigma\rangle
^2}{3g_4}\, .
\end{eqnarray}
We thus predict the existence of substructures near these
transitions, when considering fermions in the adjoint
representation. Searching for such hidden behaviors in lattice
simulations would help to further understand the nature of phase
transitions in QCD.

We have shown how deconfinement (i.e. a rise in the Polyakov loop)
is a consequence of chiral symmetry restoration in the presence of
fermions in the fundamental presentation. In nature quarks have
small, but nonzero masses, which makes chiral symmetry only
approximate. Nevertheless, the picture presented here still holds:
confinement is driven by the dynamics of the chiral transition.
The argument can be extended even further: If quark masses were
very large then chiral symmetry would be badly broken, and could
not be used to characterize the phase transition. But in such a
case the $Z_2$ symmetry becomes more exact, and by reversing the
roles of the protagonists in the previous discussion, we would
find that the $Z_2$ breaking drives the (approximate) restoration
of chiral symmetry. Which of the underlying symmetries demands and
which amends can be determined directly from the critical behavior
of the spatial correlators of hadrons or of the Polyakov loop
\cite{Mocsy:2003tr}.

With quarks in the adjoint representation we investigated the
physical scenario \cite{Karsch:1998qj} in chiral symmetry is
restored after deconfinement set in, i.e. $T_{\rm{c}\chi}\ll
T_{\rm{c}\sigma}$. In this case we have pointed to the existence
of an interesting structure, which was hidden until now: There are
still two distinct phase transitions, but since the fields are now
entangled, the transitions are not independent. This entanglement
is shown at the level of expectation values and spatial
correlators of the fields.
 Lattice simulations will play an important role in checking these predictions.

The analysis can be extended for phase transitions driven by a
chemical potential. {}For 2 colors there is a phase transition
from a quark-antiquark condensate to a diquark condensate
\cite{Hands:2001jn}. We hence predict, in two color QCD, that when
diquarks form for $\mu=m_{\pi}$, the Polyakov loop also feels the
presence of the phase transition exactly in the same manner as it
feels when considering the temperature driven phase transition.
Such a situation is supported by recent lattice simulations
\cite{Alles:2002st}. The results presented here are not limited to
describing the chiral/deconfining phase transition and can readily
be used to understand phase transitions sharing similar features.


\end{document}